\documentstyle[sprocl]{article}
\def\Journal#1#2#3#4{{#1} {\bf #2}, #3 (#4)}

\def\PLB{{\em Phys. Lett.}  B}

\def\ARNPS{\em Ann. Rev. Nucl. Part. Sci.}

\newcommand{\dphi}{\mbox{$\delta \phi$}}
\def\sm{SM}

\def\gsim{{~\raise.15em\hbox{$>$}\kern-.85em
          \lower.35em\hbox{$\sim$}~}}
\def\lsim{{~\raise.15em\hbox{$<$}\kern-.85em
          \lower.35em\hbox{$\sim$}~}}

\def\ie{{\it i.e.}}
\def\eg{{\it e.g.}}
\def\etal{{\it et al.}}

\def\beq{\begin{equation}}
\def\eeq{\end{equation}}
\def\beqa{\begin{eqnarray}}
\def\eeqa{\end{eqnarray}}

\def\sm{Standard Model}

\begin{document}
\rightline{SLAC-PUB-7568}
\rightline{July 1997}
\medskip
\renewcommand{\thefootnote}{\fnsymbol{footnote}}
\footnotetext{Presented by Mihir P. Worah. Research supported by the
department of Energy under contract DE-AC03-76SF00515.}
\title{$CP$ VIOLATION DUE TO NEW $\Delta B=1$ AMPLITUDES}
\author{Y. GROSSMAN and M. P. WORAH}
\address{Stanford Linear Accelerator Center \\
        Stanford University, Stanford, CA 94309}

\maketitle\abstracts{
We make a systematic analysis of the effects of new physics in
the $B$ decay amplitudes on the $CP$ asymmetries in 
neutral $B$ decays.
Although these are expected to be smaller than new physics effects 
on the mixing 
amplitude, they are easier to probe in some cases.
The effects of new contributions to the mixing amplitude are felt 
universally across all decay modes, whereas the effects of new decay 
amplitudes could vary from mode to mode.
In particular the prediction that 
the $CP$ asymmetries in the $B$ decay modes with
$b\to c\bar c s$, $b \to c\bar c d$, $ b\to c \bar u d$ 
and $ b\to s \bar s s$ 
should all measure the same quantity ($\sin2\beta$ in the \sm)
could be violated. 
}
\section{Introduction}

In the limit of one dominant decay amplitude,
the $CP$ violating asymmetries measured in the time dependent
decays of neutral $B$
mesons to $CP$ eigenstates depend only on the sum of the phase of the 
$B^0 - \bar B^0$ mixing amplitude and the phase of the decay 
amplitude.\cite{Yossi}
The size of these phases, however, is at present very much uncertain.
Thus, the currently allowed range for the $CP$ asymmetries measurements in 
$B$ decays is very large. Yet, there exists a 
precise prediction concerning the
$CP$ asymmetries in $B$ decays made by the Standard Model:$\;$\cite{us}
\begin{itemize}
\item{The $CP$ asymmetries in all $B$ 
decays that do not involve direct $b \to u$ (or $b \to d$)
transitions have to be the same.}
\end{itemize}

New physics could in principle contribute to both the mixing matrix and
to the decay amplitudes. 
The distinguishing feature of new
physics in mixing matrices is that its effect is universal, \ie,
although it changes the magnitude of the asymmetries it
does not change the patterns predicted by the \sm.
In particular, the above prediction is not violated. 
Thus, the best way to search for these effects
would be to compare the observed $CP$ asymmetry in a particular
decay mode with the asymmetry predicted in the \sm. 
However, due to the uncertainties in the \sm\ predictions
these effects have to be large, and even then may be
hard to detect.\cite{gnw}

In contrast, the effects of new physics in decay amplitudes are
manifestly non-universal, \ie,
they depend on the specific process and
decay channel under consideration. Experiments 
on different decay modes that would measure
the same $CP$ violating quantity in the absence of new contributions
to decay amplitudes, measure in this case different $CP$ violating
quantities. Thus, the above mentioned prediction
can be violated.\cite{us}

\section{The Effects of New Decay Amplitudes}


Consider $B$ decay into $CP$ eigenstates
where the decay
amplitude $A$ contains contributions from two terms with 
magnitudes $A_i$,
$CP$ violating phases $\phi_i$ and $CP$ conserving phases $\delta_i$
(in what follows it will be convenient to think of $A_1$ giving the
dominant \sm\ contribution, and $A_2$ giving the sub leading \sm\
contribution or the new physics contribution)
\beq
A = A_1e^{i\phi_1}e^{i\delta_1} + A_2e^{i\phi_2}e^{i\delta_2}, \qquad
\bar A = A_1e^{-i\phi_1}e^{i\delta_1} + A_2e^{-i\phi_2}e^{i\delta_2}.
\eeq
To first order in $r \equiv A_2/A_1$ 
the time dependent $CP$ asymmetry for the decays of states that
were tagged as pure $B^0$ or $\bar B^0$ at production into $CP$
eigenstates is given by \cite{Gronau}
\beq \label{acp}
a_f(t) = a_f^{\cos} \cos(\Delta Mt) + a_f^{\sin} \sin(\Delta Mt),
\eeq
with
\beqa
a_f^{\cos} &=& -2r\sin(\phi_{12})\sin(\delta_{12}), \nonumber \\
a_f^{\sin}&=& \sin 2(\phi_M + \phi_1) + 2r\cos 2(\phi_M + \phi_1)
\sin(\phi_{12})\cos(\delta_{12}),
\label{a2}
\eeqa
and we have defined 
$\phi_{12}=\phi_1-\phi_2$ and $\delta_{12}=\delta_1-\delta_2$.

In the case $r=0$ or $\phi_{12}=0$ one recovers the frequently 
studied case where $a_f^{\cos}=0$ and $a_f^{\sin} = 
-\sin 2(\phi_M + \phi_1)$.
In decay modes where the above condition
holds (to a good approximation) in the \sm, one is sensitive to new physics
if it results in $r \ne 0$ and $\phi_{12}\ne 0$. There are
several ways one can look for this kind of new physics \eg:

$(a)$ Direct $CP$ violation. This occurs when
$\delta_{12} \ne 0$ and can be measured by a careful
study of the time dependence since it gives rise to $a_f^{\cos} \ne 0$.
Such a scenario would
also give rise to $CP$ asymmetries in charged $B$ decays.
While this would be a clear
signal of new physics in the decay amplitude, the opposite is not true.
In cases where the relative strong phase between the new physics and the 
\sm\ amplitudes vanishes, the \sm\ predictions continue to hold.

$(b)$ Different quark level 
decay channels that measure the same phase when only one
amplitude contributes, now measure different phases if more than one
amplitude contributes, \ie\ two different processes with the same
$\phi_1$, but with different $r$ or $\phi_2$.
This does not depend on new strong phases, and we concentrate on this.

To this end we write
\beq
a_f^{\sin}=-\sin 2(\phi_0 + \dphi), \label{aa}
\eeq
where $\phi_0$ is the phase predicted at leading order in the \sm, and
$\dphi$ is the correction to it. 
The cleanest \sm\ predictions are found in the modes that measure
$\beta$ when $r=0$, and we concentrate on these.
These are $b \to  c \bar c s$ (\eg\ $B \to \psi K_S$), 
$b \to  c \bar c d$ 
(\eg\ $B \to D^{+}D^{-}$),
$b \to c \bar u d$ 
(\eg\ $B \to D_{CP}\rho$) and
$b \to s \bar s s$,\cite{london} (\eg\
$B \to \phi K_S$).
We now estimate the size of the sub-leading
\sm\ corrections to the above processes, which then allows us to
quantify how large the new physics effects have to be in order for
them to be probed.

There is a \sm\ penguin contribution to $B \to \psi K_S$.
However, as is well known, this contribution has the same phase as the
tree level contribution and
hence $\dphi_{SM}=0$ in Eq.~(\ref{aa}). 
The mode $b \to c \bar c d$ also has a penguin correction in the \sm. 
However,  in this case 
$\phi_{12}={\cal O}(1)$ and we estimate
the correction as 
\beq
\dphi_{SM}(B \to D^+D^-) \simeq
\frac{V_{tb}V_{td}^*}{V_{cb}V_{cd}^*}\frac{\alpha_s(m_b)}{12 \pi}
\log(m_b^2/m_t^2)\lsim 0.1 ,
\eeq
where the upper bound is obtained for $|V_{td}| < 0.02$, $m_t=180$ GeV
and $\alpha_s(m_b)=0.2$. 
Recent estimates \cite{Marti} found that an even larger value of up to
$\dphi_{SM} \sim 0.3$ cannot be excluded. 
The mode $b \to c \bar u d$ does not get
penguin corrections, however there is a doubly Cabbibo suppressed tree
level correction coming from $b \to u\bar c d$. Thus $B \to
D_{CP} \rho$ gets a second contribution with different CKM elements. 
We estimate
\beq
\dphi_{SM}(B \to D_{CP}\rho) =
\frac{V_{ub}V_{cd}^{*}}{V_{cb}V_{ud}^{*}} r_{FA}
\le 0.05 .
\eeq
where $r_{FA}$ is the ratio of matrix elements with $r_{FA}=1$ in the
factorization approximation.
We have used $|V_{ub}/V_{cb}| < 0.11$, and what we believe is a 
reasonable limit for the matrix elements ratio,
$r_{FA}<2$, to obtain the upper bound. 

For the neutral current modes $B \to \phi K_S$ 
we use CKM unitarity to 
write the decay amplitude as a sum of two terms \cite{gq}
\beq \label{twoterms-s}
A = V_{cb}V_{cs}^* A^{ccs} + V_{ub}V_{us}^* A^{uus}.
\eeq
The correction due to the second term is doubly Cabbibo suppressued.
Assuming that $\phi$ is a pure $s\bar s$ state and
for a very heavy $b$ quark $A^{uus}/A^{ccs} = 1$. $A^{uus}$ may be enhanced
due to the finite $b$ mass, 
or due to $SU(3)_{flavor}$ mixing
since the $\phi$ also contains a small $u \bar u$ component.
The first effect is unlikely to be large, and the second is
experimentally known to be small.
Thus, we believe that a  
reasonable limit for the matrix elements ratio is $A^{uus} /A^{ccs} < 2$,
leading to
\beq
\dphi_{SM}(b \to s \bar ss) \le 0.05.
\eeq

We emphasize the importance of the decay 
$B \to \phi K_S$ in probing new physics. 
Since this is a penguin mediated decay in the \sm, it is the most
sensitive of all the modes we addressed to the possible contributions
from new physics. Moreover, although technically classified as a
``rare decay'' the $CP$ asymmetry in this mode is possibly 
measurable at the early stages of the $B$ factories. This is because
the recent CLEO measurements of the $B \to \pi K$ and $B \to \eta ' K$
branching ratios suggest that penguin induced decays are large,
and one can then estimate $BR(B_D \to \phi K_S) \sim 10^{-5}$.
An important
advantage is that the efficiency for tagging this mode is very large.
One can reconstruct the $\phi$ using the decay $\phi \to K^+
K^-$ which has a large branching ratio $\sim 50\%$ and very high
efficiency (probably up to $80-90\%$). 
The other two dominant decay modes $\phi \to K_LK_S$ and 
$\phi \to \pi\rho$ can also be used (probably with smaller efficiency.)
Thus, 
the effective rate for 
$B \to \phi K_S$ is within a factor of ten of $B \to \psi K_S$ and
of the same order as $B \to \pi^+\pi^-$.


\begin{table}
\caption[tbmodels]
{Summary of the useful modes. The ``SM angle'' entry corresponds to
the angle obtained 
assuming one decay amplitude. 
New contributions to the mixing amplitude would shift all the entries
by $\delta_{m_d}$.
$\dphi$ (defined in Eq.~(\ref{aa}))
corresponds to the (absolute value of the) correction to the universality
prediction within each model:
$\dphi_{SM}$ -- \sm, $\dphi_{A}$ -- Effective Supersymmetry, 
$\dphi_{B}$ -- Models with Enhanced Chromomagnetic Dipole Operators and
$\dphi_{C}$ -- Supersymmetry without R-parity. 
$1$ means that the phase can get any value.
}
\vspace{0.4cm} 
 \begin{tabular}{|c|c|c|c|c|c|c|}
\hline 
& & & & & & \\
\qquad Mode \qquad &  SM angle $(\phi_0)$ & $\dphi_{SM}$ & $ \dphi_A$ & 
$\dphi_B$ & $\dphi_C$ & $BR$\\
\hline 
$B \to \psi K_S$ & $\beta$ & 0 & $ 0.1 $ & $ 0.1 $ & $ 0.1 $ &
 $7 \times 10^{-4}$
\\ 
$B \to D^+D^-$ & $\beta$ & $ 0.1$ & $ 0.2$ &$ 0.6$ &$ 0.6$ & 
 $4 \times 10^{-4}$ \\
$B \to D_{CP} \rho$ & $\beta$ & $ 0.05$ & 0 & 0 & $ 0.5$ &
 $10^{-5}$ \\
$B \to \phi K_S$ & $ \beta $ & 0.05 & $ 1$ & $ 1$ & 0 &
 $10^{-5}$ \\
\hline
\end{tabular}
\label{sumtab2}
\end{table}

We have studied three models:$\;$\cite{us}
(a) Effective Supersymmetry, (b) Models with enhanced Chromomagnetic
dipole operators, and (c) Supersymmetry without R parity.
In Table 1 we show the largest
allowable effects in these models. 
Such effects in general supersymmetric models
were also studied.\cite{Mart-Barb} 
In all of the examples one finds that large experimentally detectable
effects are possible.

\section{Conclusions}
$CP$ asymmetries in $B$ decays can be a very useful
tool in looking for physics beyond the \sm.
One can look for new contributions to the $B-\bar B$ mixing. However,
these have to be large to be discovered. Alternatively, new
contributions to the $B$ decay amplitudes can be discovered even if
they are rather small, $O(5\%)$.
With a large strong phase, this new contribution
can lead to sizeable direct $CP$ violation. 
Even without it, such effects can be probed 
by comparing two experiments that measure the same phase $\phi_0$ in
the \sm\ [see Eq. (\ref{aa})].
The most promising way to look for new physics effects in decay
amplitudes is to compare all
the $B$ decay modes that measure $\beta$ in the \sm.
The best mode is $B \to \psi K_S$ which has a sizeable rate and negligible 
theoretical uncertainty. This mode should be the reference mode to which all
other measurements are compared.
The $b \to c \bar u d$ and $b \to s \bar s s$ modes are also 
theoretically very clean.
Thus, the two ``gold plated'' relations are 
\beq
|\phi(B \rightarrow \psi K_S) - \phi(B \rightarrow \phi K_S)| < 0.05,
\eeq
and
\beq
|\phi(B \rightarrow \psi K_S) - \phi(B \rightarrow D_{CP} \rho)| < 0.05. 
\eeq
Any deviation from these two relations will be a clear indication for new 
physics in decay amplitudes. The mode $B \to \phi K_S$ is particularly 
sensitive to new physics since it is a loop induced rare decay in the \sm.

\end{document}